\documentclass[algorithms,article,accept,moreauthors,pdftex,10pt,a4paper]{mdpi} 
\firstpage{1} 
\articlenumber{x}
\doinum{10.3390/------}
\pubvolume{xx}
\pubyear{2017}
\copyrightyear{2017}
\history{}
\pdfoutput=1
\fancypagestyle{plain}{}
\makeatletter
\renewcommand{\@maketitle}{
	\ifthenelse{\equal{\@arttype}{Supfile}}{%
	\begin{flushleft}
		\fontsize{18}{18}\selectfont
		\raggedright
		\noindent\textbf{Supplementary Materials: \@Title}%
		\par
		\vspace{12pt}
		\fontsize{10}{10}\selectfont
		\noindent\boldmath\bfseries{\@Author}
	\end{flushleft}
		}{%
		\begin{flushleft}
		\ifthenelse{\equal{\@arttype}{Book}}{%
			}{%
			\vspace*{-1.75cm}
			}
		{
		\ifthenelse{\equal{\@arttype}{Preprints} \OR \equal{\@arttype}{Book}}{%
			}{%
			\ifthenelse{\equal{\@arttype}{Conference Proceedings Paper}}{%
				\includegraphics[height=1.2cm]{logo-conference}%
				\hfill \href{http://www.mdpi.com}{\includegraphics[height=1cm]{logo-preforum}}%
				}{%
				\ifthenelse{\equal{\@status}{submit}}{%
					\hfill \href{http://www.mdpi.com}{%
					\includegraphics[height=1cm]{logo-mdpi}}\vspace{0.5cm}%
					}{
					\href{http://www.mdpi.com/journal/\@journal}{
                    }%
					\hfill \ifthenelse{\equal{\@journal}{scipharm}}{%
						\href{http://www.mdpi.com}{\includegraphics[height=1cm]{logo-mdpi-scipharm}}%
						}{%
						}%
					}%
				}%
			}%
		\par
		}
		{
    		\vspace{14pt}
    		\fontsize{10}{10}\selectfont
		\ifthenelse{\equal{\@arttype}{Book}}{
			}{
    			\ifthenelse{\equal{\@arttype}{Preprints}}{
				\textit{Article}%
				}{%
				\ifthenelse{\equal{\@arttype}{Reprint}}{%
					\textit{\ifthenelse{\equal{\@originalarttype}{\@empty}}{Article}{\@originalarttype}}%
					}{%
					}%
				}
			}%
 	   	\par%
    		}
    		{
   	 	\vspace{-1pt}
  	  	\fontsize{18}{18}\selectfont
   	 	\boldmath\bfseries{\@Title}
   	 	\par
   	 	\vspace{15pt}
   	 	}
   		{
    		\boldmath\bfseries{\@Author}
    		\par
    		\vspace{-4pt}
    		}
    		\end{flushleft}
  	  }%
	}
\makeatother
\renewcommand{\cright}{}

\newlength{\normalparindent}
\AtBeginDocument{\setlength{\normalparindent}{\parindent}}

\usepackage{siunitx}
\usepackage{amssymb}
\usepackage[a]{esvect}
\usepackage{subfig}
\usepackage{float}
\usepackage{listings}
\usepackage{dcolumn}
\newcolumntype{d}[1]{D{.}{\cdot}{#1} }
\newcolumntype{Y}{D{.}{.}{1.2}}

\usepackage{algorithm}
\usepackage[noend]{algpseudocode}
\algnewcommand{\algorithmicand}{\textbf{ and }}
\algnewcommand{\algorithmicor}{\textbf{ or }}
\algnewcommand{\OR}{\algorithmicor}
\algnewcommand{\AND}{\algorithmicand}
\algnewcommand{\var}{\texttt}
\usepackage{setspace}

\algrenewcommand\alglinenumber[1]{{\sffamily\footnotesize#1}}
\makeatletter
\algrenewcommand\ALG@beginalgorithmic{\fontsize{10}{16}\selectfont}
\makeatother

\makeatletter
\newcommand{\algmargin}{\the\ALG@thistlm}
\makeatother
\newlength{\whilewidth}
\algdef{SE}[parWHILE]{parWhile}{EndparWhile}[1]
  {\parbox[t]{\dimexpr\linewidth-\algmargin}{%
     \hangindent\whilewidth\strut\algorithmicwhile\ #1\ \algorithmicdo\strut}}{\algorithmicend\ \algorithmicwhile}%
\algnewcommand{\parState}[1]{\State%
  \parbox[t]{\dimexpr\linewidth-\algmargin}{\linespread{0.7}\selectfont\strut #1\strut}}

\usepackage{environ}
\NewEnviron{myequation}{%
\begin{equation}
\scalebox{0.9}{$\BODY$}
\end{equation}}

\Title{Mapping Higher-Order Network Flows in Memory and Multilayer Networks with Infomap}

\Author{Daniel Edler, Ludvig Bohlin, and Martin Rosvall *}

\AuthorNames{Daniel Edler, Ludvig Bohlin, and Martin Rosvall}

\address[1]{\noindent Integrated Science Lab, Department of Physics, Ume{\aa} University, SE-901 87 Ume{\aa}, Sweden}

\corres{\hangafter=1 \hangindent=1.05em \hspace{-0.82em} Correspondence: martin.rosvall@umu.se}

\abstract{%
Comprehending complex systems by simplifying and highlighting important dynamical patterns requires modeling and mapping higher-order network flows. However, complex systems come in many forms and demand a range of representations, including memory and multilayer networks, which in turn call for versatile community-detection algorithms to reveal important modular regularities in the flows. Here we show that various forms of higher-order network flows can be represented in a unified way with networks that distinguish physical nodes for representing a~complex system's objects from state nodes for describing flows between the objects. Moreover, these so-called sparse memory networks allow the information-theoretic community detection method known as the map equation to identify overlapping and nested flow modules in data from a range of~different higher-order interactions such as multistep, multi-source, and temporal data. We derive the map equation applied to sparse memory networks and describe its search algorithm Infomap, which can exploit the flexibility of sparse memory networks. Together they provide a general solution to reveal overlapping modular patterns in higher-order flows through complex systems.
}


\begin{document}




\section{Introduction}

\begin{figure}[H]
    \centering
	\includegraphics[width=0.9\columnwidth]{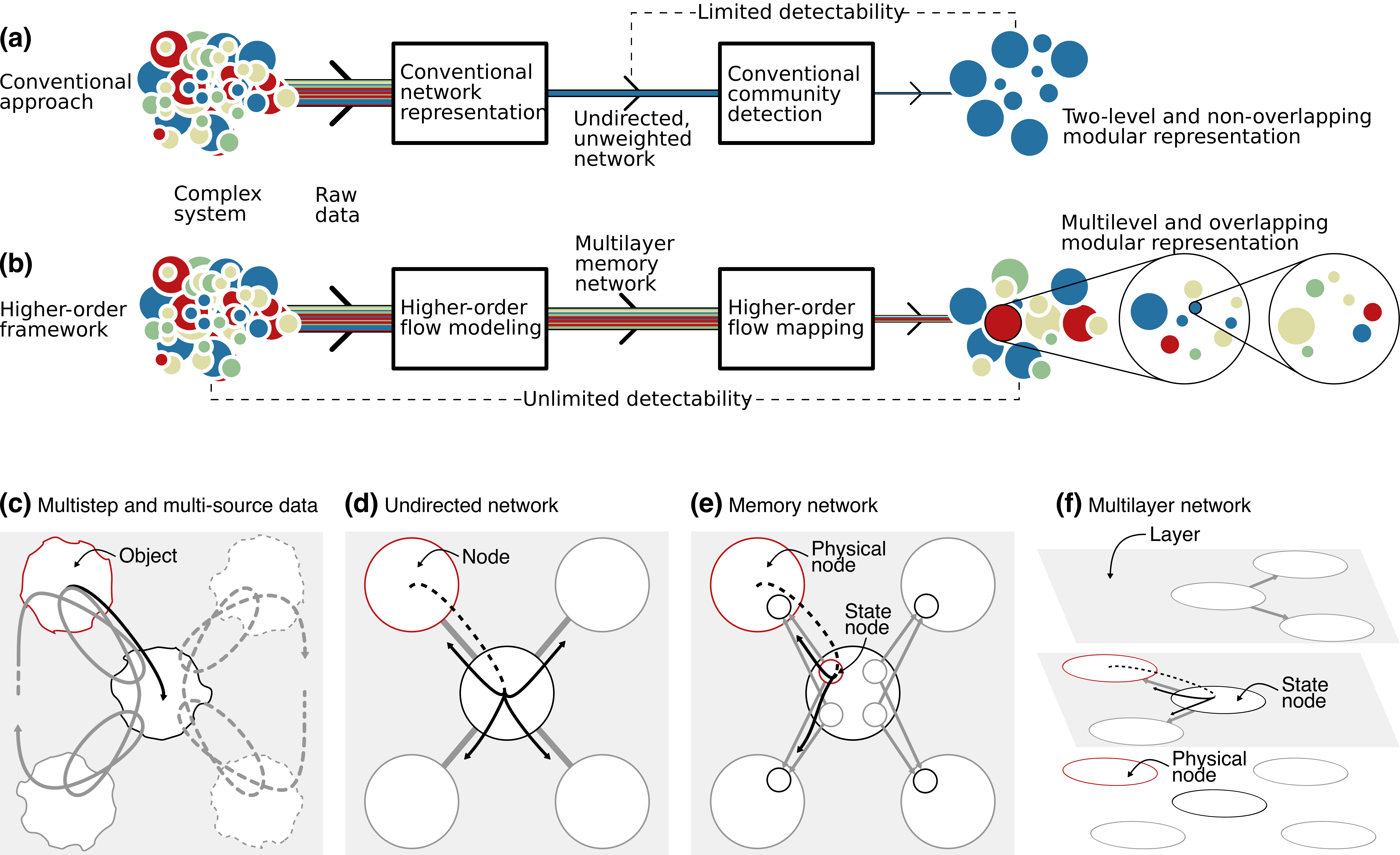}
    \caption{Going beyond standard network representations makes it possible to take advantage of richer interaction data. (\textbf{a}) Standard methods shoehorn interaction data about a complex system into an often unweighted and undirected network, which limits what regularities can be detected; (\textbf{b}) Modeling and mapping higher-order network flows can break this detectability limit; (\textbf{c}) System flows from two data sources between objects. The black arrow shows how flows can come to the center object; (\textbf{d}) The system represented as an undirected network with nodes for the objects. The links show where flows coming in to the center node are constrained to go next; (\textbf{e}) The system represented as a memory network with~physical nodes for the objects and state nodes for constraining flows along their links. The links show where flows coming in to the center node are constrained to go next depending on where they come from; (\textbf{f}) The system represented as a multilayer network with physical nodes for the objects and~state nodes in layers corresponding to different data sources. The links show where the flows coming in to the center node are constrained to go next depending on which layer they are in.\label{fig:intro}} 
\end{figure}

To connect structure and dynamics in complex systems, researchers model, for example, people navigating the web~\cite{brin1998anatomy}, rumors wandering around among citizens~\cite{newman2005measure}, and passengers traveling through~airports~\cite{lordan2015study}, as flows on networks with random walkers. Take an air traffic network as~an~example. In a standard network approach, nodes represent airports, links represent flights, and~random walkers moving on the links between the nodes represent passengers. This dynamical process corresponds to a first-order Markov model of network flows: a passenger arriving in~an~airport will randomly continue to an airport proportional to the air traffic volume to that airport. That~means, for example, that two passengers who arrive in Las Vegas, one from San Francisco and one from~New York, will have the same probability to fly to New York next. In reality, however, passengers are more likely to return to where they come from~\cite{rosvall2014memory}. Accordingly, describing network flows with a first-order Markov model suffers from memory loss and washes out significant dynamical patterns~\cite{belik2011natural,pfitzner2013betweenness,poletto2013human} (Figure~\ref{fig:intro}a). Similarly, aggregating flow pathways from multiple sources, such as different airlines or~seasons in the air traffic example, into a single network can distort both the topology of~the~network and the dynamics on the network~\cite{mucha2010community,kivela2013multilayer,boccaletti2014structure,de2015identifying} (Figure~\ref{fig:intro}c,d).

As a consequence, the actual patterns in the data, such as pervasively overlapping modules in air traffic and social networks~\cite{ahn2010link}, cannot be identified with community-detection algorithms that operate on first-order network flows~\cite{rosvall2014memory}. To take advantage of higher-order network flows, researchers have therefore developed different representations, models, and community-detection algorithms that broadly fall into two research topics: memory networks and multilayer networks. In memory networks, higher-order network flows can represent multistep pathways such as flight itineraries~\cite{rosvall2014memory,peixoto2015modeling,xu2016representing,scholtes2017network}, and~in~multilayer networks, they can represent temporal or multi-source data such as multiseason air traffic~\cite{mucha2010community,de2013mathematical,de2015identifying,wehmuth2016multiaspect} (Figure~\ref{fig:intro}b). Whereas memory networks can capture where flows move to~depending on where they come from (Figure~\ref{fig:intro}e), multilayer networks can capture where flows move to~depending on the current layer (Figure~\ref{fig:intro}f).
Because memory and multilayer networks originate from different research topics in network science, revealing their flow modules have been considered as disparate community detection problems. 
With the broad spectrum of higher-order network representations required to best describe diverse complex systems, the apparent need for~specialized community-detection methods raises the question: How can we reveal communities in higher-order network flows with a general approach?

Here we show that describing higher-order network flows with so-called sparse memory networks and~identifying modules with long flow persistence times, such as groups of airports that contain frequently traveled routes, provides a general solution to reveal modular patterns in higher-order flows through complex systems. When modeling flows in conventional networks, a single node type both represents a complex system's objects and describes the flows with the nodes' links (Figure~\ref{fig:intro}c,d). Sparse memory networks, however, discriminate physical nodes, which represent a complex system's objects, from state nodes, which describe a complex system's internal flows with~their links (Figure~\ref{fig:intro}e,f). In~sparse memory networks, state nodes are not bound to represent, for~example, previous steps in memory networks or layers in multilayer networks, but are free to represent abstract states such as lumped states~\cite{persson2016maps} or states in multilayer memory networks, which we demonstrate with multistep and~multiquarter air traffic data. In this way, a sparse memory network is a MultiAspect Graph with two aspects: the physical object and the flow state such as memory or~layer~\cite{wehmuth2016multiaspect}. We show that various higher-order network flow representations, including memory and~multilayer networks, can be represented with sparse memory networks. We also provide a detailed derivation of~the~information-theoretic map equation for identifying hierarchically nested modules with long flow persistence times in sparse memory networks, and introduce a new version of the map equation's search algorithm Infomap that exploits the flexibility of sparse memory networks.  

\section{Modeling Network Flows}

The dynamics and function of complex systems emerge from interdependences between their components~\cite{barabasi1999emergence,barrat2004architecture,boccaletti2006complex}. Depending on the system under study, the components can be, for example, people, airports, hospital wards, and banks. The interdependence, in turn, often~comes from flows of some entity between the components, such as ideas circulating among colleagues, passengers traveling through airports, patients moving between hospital wards, or money transferred between banks. To efficiently capture such flows through complex systems, researchers model them with random walks on networks.

\subsection{First-Order Network Flows}

In a first-order network representation, the flow direction only depends on the previously visited physical node. That is, for a random walker that moves between physical nodes $i \in \{1,2,\ldots,N_P\}$, which represent objects in a complex system, and in $t$ steps generates a random variable sequence $X_1,X_2,\ldots,X_t$, the transition probabilities
\begin{equation}
\label{eq1}
P(X_t | X_{t-1}, X_{t-2}, \ldots ) = P(X_t | X_{t-1})
\end{equation}
only depend on the previously visited node's outlinks.
With link weights $w_{ij}$ between physical nodes $i$ and $j$, and total outlink weights $w_i = \sum_{j} w_{ij}$ of node $i$, the first-order transition probabilities are
\begin{equation}
\label{eq2}
P(i \rightarrow j) = P_{ij} = \frac{w_{ij}}{w_i}, 
\end{equation}
which give the stationary visit rates
\begin{equation}
\label{eq3}
\pi_i = \sum_j \pi_jP_{ji}.
\end{equation}

To ensure ergodic stationary visit rates, from each node we can let the random walker teleport with~probability $\tau$, or with probability $1$ if the node has no outlinks, to a random target node proportional to the target node's inlink weight~\cite{lambiotte2012ranking}. 

While a first-order model is sufficient to capture flow dynamics in some systems, recent studies have shown that higher-order flows are required to capture important dynamics in many complex systems \cite{song2010limits,belik2011natural,pfitzner2013betweenness}. The standard approach of modeling dynamical processes on networks with~first-order flows oversimplifies the real dynamics and sets a limit of what can actually be detected in the system (Figure~\ref{fig:intro}). Capturing critical phenomena in the dynamics and~function of complex systems therefore often requires models of higher-order network flows~\cite{meiss2008ranking,chierichetti2012web,singer2014markovnav,takaguchi2011predictability,holme2012temporal,pfitzner2013betweenness,song2010limits,belik2011natural,poletto2013human}.

\subsection{Higher-Order Network Flows}
In higher-order network representations, more information than the previously visited physical node’s outlinks are used to determine where flows go. Examples of higher-order network representations are memory, multilayer, and temporal networks.

In memory networks, the flow direction depends on multiple previously visited nodes. Specifically, for a random walker that steps between physical nodes and generates a sequence of~random variables $X_1,\ldots,X_t$, the transition probabilities for a higher-order flow model of order $m$,
\begin{equation}
\begin{split}
P(X_t | X_{t-1}, X_{t-2}, \ldots ) = P(X_t | X_{t-1}, \ldots, X_{t-m}),
\label{eq:homarkov}
\end{split}
\end{equation}
depend on the $m$ previously visited physical nodes. Assuming stationarity and $m$ visited nodes from~$x_{-m}$ $m$ steps ago to previously visited $i$ in sequence $\vv{x_{-m}x_{-m+1}\cdots x_{-2}i}$, the $m$th-order transition probabilities between physical nodes $i$ and $j$ correspond to first-order transition probabilities $P(\vv{x_{-m}\cdots x_{-2}i} \to \vv{x_{-m+1}\cdots x_{-2}ij})$ between state nodes $\alpha_i = \vv{x_{-m}\cdots x_{-2}i}$ and $\beta_j = \vv{x_{-m+1}\cdots x_{-2}ij}$. We use subscript $i$ in $\alpha_i$ to highlight the state node's physical node. With link weights $w_{\alpha_i\beta_j}$ between~state nodes $\alpha_i$ and $\beta_j$, and total outlink weights $w_{\alpha_i} = \sum_{\beta_j} w_{\alpha_i \beta_j}$, the transition probabilities for memory networks are
\begin{equation}
P(\vv{x_{-m}\cdots x_{-1}i} \to \vv{x_{-m+1}\cdots x_{-1}ij}) = P_{\alpha_i\beta_j} = \frac{w_{\alpha_i\beta_j}}{w_{\alpha_i}}.\label{eq:memorytransition}
\end{equation}

As in a first-order network representation, teleportation can ensure ergodic stationary visit rates $\pi_{\alpha_i}$~\cite{rosvall2014memory}.

In multilayer networks with the same physical nodes possibly present in multiple layers, the flow direction depends on both the previously visited physical node $i$ and layer $\alpha \in \{1,2,\ldots,l\}$. Similar to~the memory representation, the multilayer transition probabilities between physical nodes $i$ in~layer $\alpha$ and $j$ in layer $\beta$ correspond to first-order transition probabilities between states nodes $\alpha_i$ and~$\beta_j$.

In many cases, empirical interaction data only contain links within layers such that $w_{\alpha_i\beta_j} \ne 0$ only for $\alpha=\beta$. Without data with links between layers, it is useful to couple layers by allowing a random walker to move between layers at a relax rate $r$. With probability $1-r$, a random walker follows a link of the physical node in the currently visited layer, and with probability $r$ the random walker relaxes the~layer constraint and follows any link of the physical node in any layer. In both cases, the random walker follows a link proportional to its weight. With total outlink weight $w_{\alpha_i} = \sum_{\beta_j} w_{\alpha_i\beta_j}$ from physical node $i$ in layer $\alpha$, and total outlink weight $w_{i} = \sum_{\alpha,\beta_j} w_{\alpha_i\beta_j}$ from physical node $i$ across all layers, which both correspond to total intralayer outlink weights as long as there are only empirical links within layers, the transition probabilities for multilayer networks with modeled interlayer transitions are
\begin{equation}
P\left({\alpha}_{i} \to {\beta}_{j}\right)(r) = P_{\alpha_i\beta_j}(r) = (1-r)\delta_{\alpha\beta}\frac{w_{\alpha_i\beta_j}}{w_{\alpha_i}}+r\frac{w_{\beta_i\beta_j}}{w_i},\label{eq:interlayer_transition}
\end{equation} 
where $\delta_{\alpha\beta}$ is the Kronecker delta, which is 1 if $\alpha=\beta$ and 0 otherwise.
In this way, the transition probabilities capture completely separated layers for relax rate $r = 0$ and completely aggregated layers for relax rate $r = 1$. The system under study and problem at hand should determine the~relax rate. In practice, for relax rates in a wide range around 0.25, results are robust in many real multilayer~networks~\cite{de2015identifying}.

With both intralayer and interlayer links such that $w_{\alpha_i\beta_j} \ne 0$ also for $\alpha \ne \beta$, either from empirical data or modeled interlayer links $P_{\alpha_i\beta_j}(r) \to w_{\alpha_i\beta_j}$, the transition probabilities for multilayer networks can be written in their most general form,
\begin{equation}
P\left({\alpha}_{i} \to {\beta}_{j}\right) = P_{\alpha_i\beta_j} = \frac{w_{\alpha_i\beta_j}}{w_{\alpha_i}},\label{eq:multilayertransition}
\end{equation}
where $w_{\alpha_i}$ again is the total outlink weight from node $i$ in layer $\beta$.
For directed multilayer networks, teleportation can ensure ergodic stationary visit rates $\pi_{\alpha_i}$.

\subsection{Sparse Memory Networks}
While memory networks and multilayer networks operate on different higher-order interaction data, the resemblance between Equations~(\ref{eq:memorytransition}) and (\ref{eq:multilayertransition}) suggests that they are two examples of a more general network representation. In a memory network model, a random walker steps between physical nodes such that the next step depends on the previous steps (Equation~(\ref{eq:memorytransition})). In a multilayer network model, instead the next step depends on the visited layer (Equation~(\ref{eq:multilayertransition})). However, as Equations~(\ref{eq:memorytransition}) and~(\ref{eq:multilayertransition}) show, both models correspond to first-order transitions between state nodes $\alpha_i$ and $\beta_j$ associated with~physical nodes $i$ and $j$ in the network,
\begin{equation}
P_{\alpha_i\beta_j} = \frac{w_{\alpha_i\beta_j}}{w_{\alpha_i}}.\label{eq:generaltransition}
\end{equation}

Consequently, the state node visit rates are
\begin{equation}
\pi_{\alpha_i} = \sum_{\beta_j} \pi_{\beta_j}P_{\beta_j\alpha_i},
\end{equation}
where the sum is over all state nodes in all physical nodes.
The physical node visit rates are
\begin{equation}
\pi_{i} = \sum_{\beta_j \in i} \pi_{\beta_j},
\end{equation}
where the sum is over all state nodes in physical node $i$. Both memory and multilayer networks can be represented with a network of physical nodes and state nodes that neither are bound to represent previous steps nor current layer. Because state nodes are free to represent abstract states, and redundant state nodes can be lumped together, we call this representation a sparse memory network. 

\subsection{Representing Memory and Multilayer Networks with Sparse Memory Network }
\begin{figure}[H]
    \centering
	\includegraphics[width=0.95\textwidth]{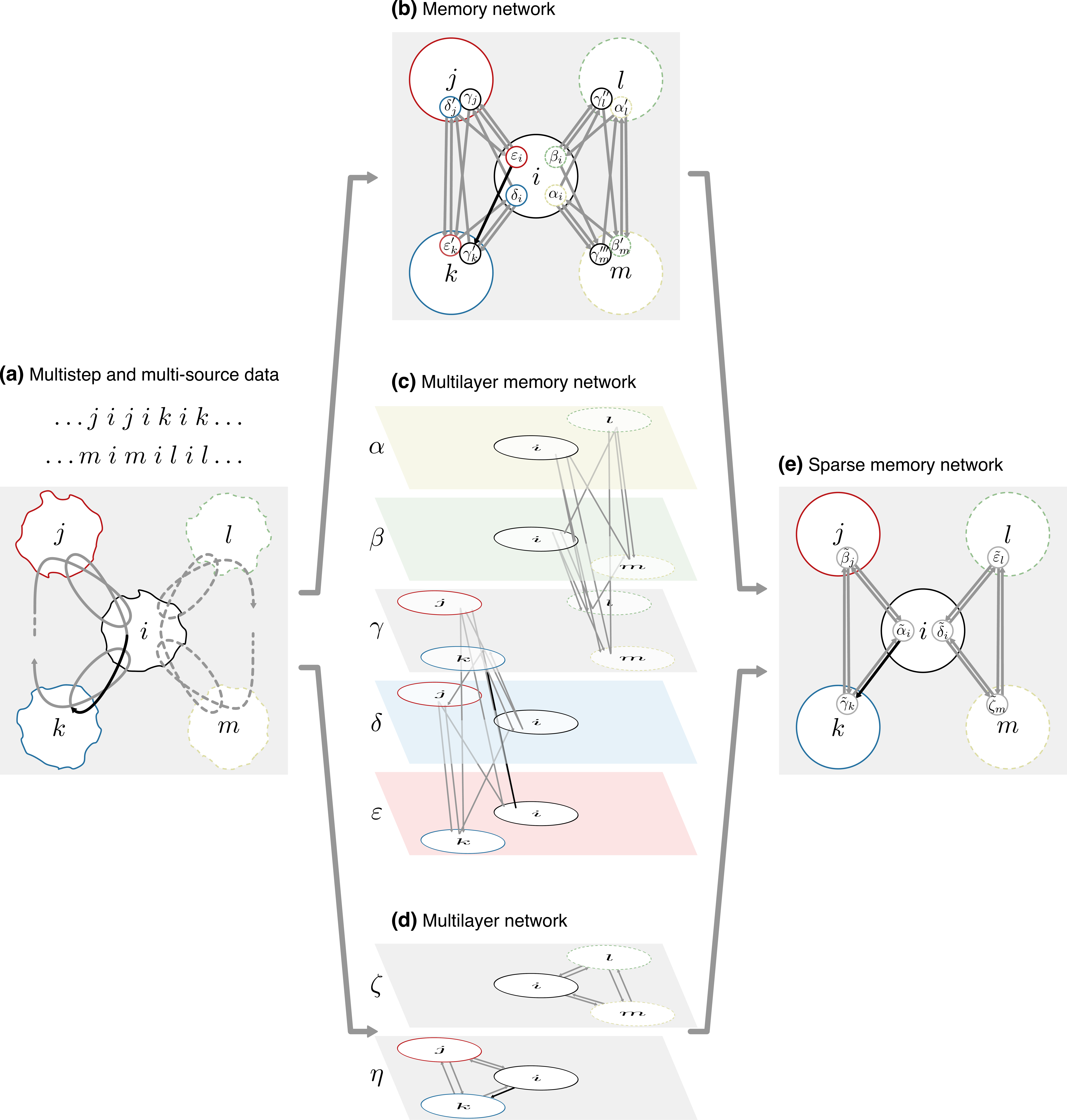}
    \caption{Modeling higher-order network flows with sparse memory networks. (\textbf{a}) Multistep pathway data from two sources illustrated on a network with five physical nodes; (\textbf{b}) The pathway data modeled with a second-order Markov model on a memory network, where state nodes capture where flows come from; (\textbf{c}) The memory network represented as a multilayer network where layers, one for each physical node, capture where flows come from; (\textbf{d}) The pathway data modeled on a two-layer network, one layer for each data source; (\textbf{e}) Both memory and multilayer networks mapped on a sparse memory network with no redundant state nodes. The black link highlights the same step in all representations. See also the dynamic storyboard available on \url{www.mapequation.org/apps/sparse-memory-network}.\label{fig:memory_representation}}
\end{figure}

To illustrate that sparse memory networks can represent both memory and multilayer networks, we use a schematic network with five physical nodes (Figure~\ref{fig:memory_representation}a). The network represents five individuals, with~the~center node's two friends to the left and two colleagues to the right. We imagine that the multistep pathway data come from two sources, say a Facebook conversation thread among the friends illustrated with the top sequence above the network and the solid pathway on~the~network, and an email conversation thread among the colleagues illustrated with the bottom sequence and~the~dashed pathway. For simplicity, the pathway among the friends stays among~the~friends and steps between friends with equal probability. The pathway among the colleagues behaves in~corresponding way. We first represent these pathway data with a memory and a multilayer network, and then show that both can be represented with a sparse memory network.

We can represent multistep pathways with links between state nodes in physical nodes. For~a~memory network representation of a second-order Markov model, each state node captures which physical node the flows come from. For example, the highlighted pathway step in Figure~\ref{fig:memory_representation}a corresponds to a link between state node $\varepsilon_i$ in physical node $i$---capturing flows coming to physical node $i$ from physical node $j$---and state node $\gamma_k'$ in physical node $k$---capturing flows coming to~physical node $k$ from physical node $i$ (Figure~\ref{fig:memory_representation}b). In this way, random walker movements between~state nodes can capture higher-order network flows between observable physical nodes. 

We can represent multistep pathways with links between state nodes in layers. First, we can map all state nodes that represent flows coming from a specific physical node onto the same layer. For~example, we map red state nodes $\varepsilon_i$ and $\varepsilon_k'$ for flows coming from red physical node $j$ to physical nodes $i$ and $k$, respectively, in Figure~\ref{fig:memory_representation}b onto the red layer $\varepsilon$ at the bottom in Figure~\ref{fig:memory_representation}c. This mapping gives one-to-one correspondence between the memory network and the multilayer network. Therefore we call it a multilayer memory network.

An alternative and more standard multilayer representation exploits that the pathway data come from two sources and uses one layer for each data source (Figure~\ref{fig:memory_representation}d). Network flows are first-order when constrained to move within individual layers and higher-order when free to move within~and~between~layers. The highlighted pathway step in Figure~\ref{fig:memory_representation}a now corresponds to a step in~layer $\eta$ between state node ${\eta}_{i}$ and state node ${\eta}_{k}$---capturing flows remaining among friends. Again, random walker movements between state nodes can capture higher-order network flows between~observable physical nodes.

In our multistep pathway example from two sources, a full second-order model in the memory network is not required and partly redundant to describe the network flows. We can model the~same pathways with a more compact description. Specifically, we can lump together state nodes $\alpha_i$ and~$\beta_i$ in~the~same physical node $i$ if they have identical outlinks, $w_{\alpha_i\gamma_j} = w_{\beta_i\gamma_j}$ for all $\gamma_j$. The lumped state node $\tilde{\alpha}_i$ replaces $\alpha_i$ and $\beta_i$ and assembles all their inlinks and outlinks such that the transition probabilities remain the same. For example, for describing where flows move from physical node $i$, state nodes $\varepsilon_i$ and $\delta_i$ have identical outlinks---reflecting that it does not matter which friend was previously active in the conversation---as well as state nodes $\beta_i$ and $\alpha_i$---reflecting that it does not matter which colleague was previously active in the conversation. Lumping together all such redundant state nodes, such as $\varepsilon_i,\delta_i \to \tilde{\alpha}_i$ or $\beta_i,\alpha_i \to \tilde{\delta}_i$, gives the sparse memory network in~Figure~\ref{fig:memory_representation}e, where state nodes no longer are bound to capture the exact previous steps but still capture the same dynamics as the redundant full second-order model. These unbound state nodes are free to~represent abstract states, such as lumped state nodes in a second-order memory network, but~can also represent state nodes in a full second-order memory network or state nodes in a multilayer network such as ${\eta}_{i} \to \tilde{\alpha}_i$ in Figure~\ref{fig:memory_representation}d and e. For example, lumping state nodes with minimal information loss can balance under- and overfitting for efficient sparse memory networks that represent variable-order network flow models~\cite{persson2016maps}. Consequently, rather than an explosion of application and data dependent representations, sparse memory networks with physical nodes and state nodes that can represent abstract states provides an efficient solution for modeling higher-order network flows.

\section{Mapping Network Flows}\label{mapequationmechanics}
While networks and higher-order models make it possible to describe flows through complex systems, they remain highly complex even when abstracted to nodes and links. Thousands or~millions of nodes and links can bury valuable information. To reveal this information, many times it is indispensable to comprehend the organization of large complex systems by assigning nodes into~modules with community-detection algorithms. Here we show how the community-detection method known as the map equation can operate on sparse memory networks and allow for versatile mapping of network flows from multistep, multi-source, and temporal data. 

\subsection{The Map Equation for First-Order Network Flows}
When simplifying and highlighting network flows with possibly nested modules, the map equation measures how well a modular description compresses the flows~\cite{rosvall2008maps}.
Because compressing data is dual to finding regularities in the data~\cite{shannon1948mathematical}, minimizing the modular description length of~network flows is dual to finding modular regularities in the network flows.
For describing movements within and between modules, the map equation uses code books that connect node visits, module exits, and module entries with code words. To estimate the shortest average description length of each code book, the map equation takes advantage of Shannon's source coding theorem and measures the Shannon entropy of the code word use rates~\cite{shannon1948mathematical}. Moreover, the map equation uses a hierarchically nested code structure designed such that the description can be compressed if~the~network has modules in which a random walker tends to stay for a long time.
Therefore, with~a~random walker as a proxy for real flows, minimizing the map equation over all possible network clusterings reveals important modular regularities in network flows.

\begin{figure}[H]
    \centering
	\includegraphics[width=0.7\textwidth]{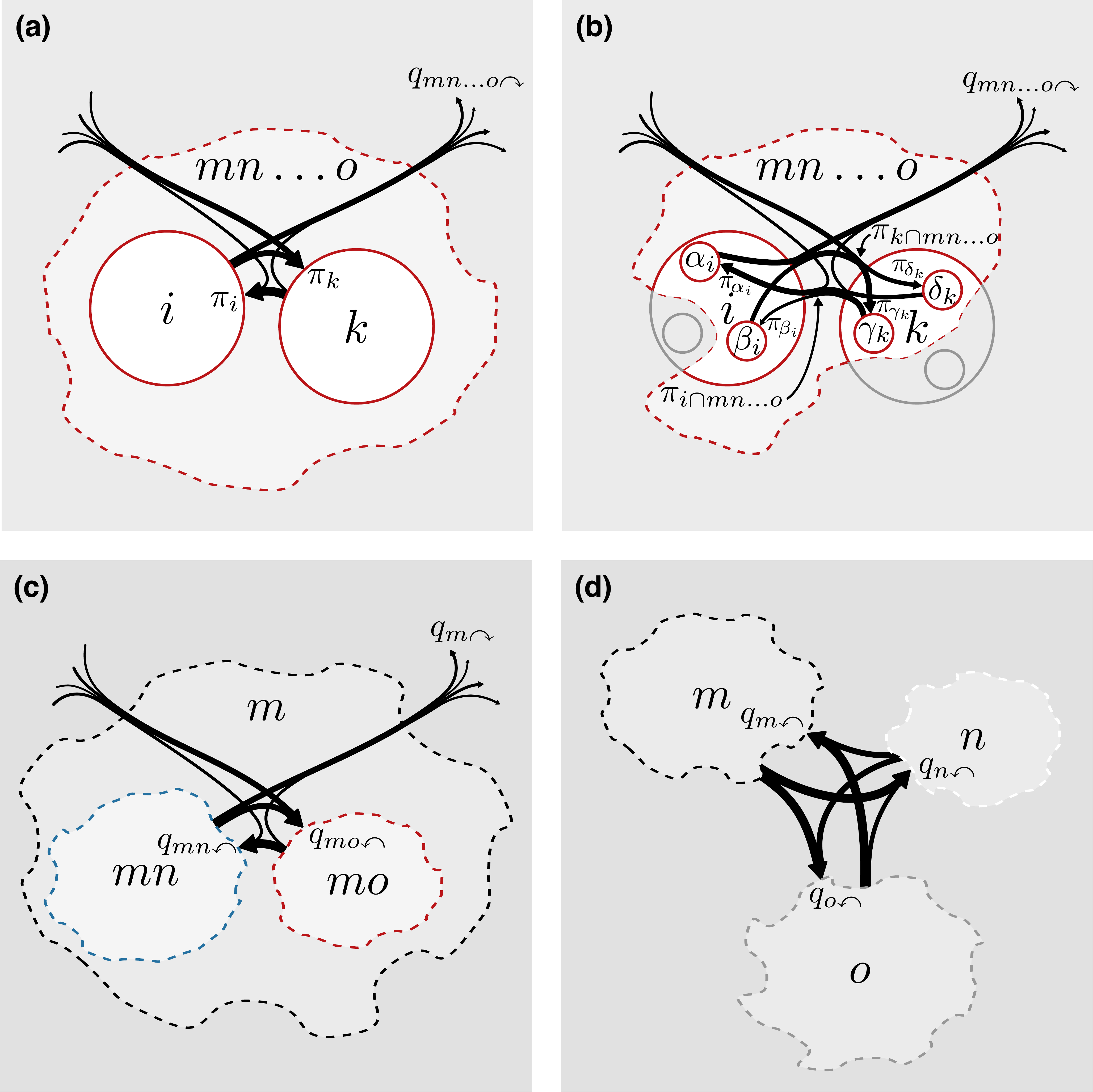}
    \caption{Network flows at different modular levels. Large circles represent physical nodes, small circles represent~state nodes, and dashed areas represent modules. (\textbf{a}) Finest modular level with physical nodes for~first-order network flows; (\textbf{b}) Finest modular level with physical nodes and state nodes for~higher-order network flows; (\textbf{c}) Intermediate level; (\textbf{d}) Coarsest modular level. \label{fig:codinglevels}}%
\end{figure}

In detail, the map equation measures the minimum average description length for a multilevel map $\mathsf{M}$ of $N$ physical nodes clustered into $M$ modules, for which each module $m$ has a submap $\mathsf{M}_m$ with $M_m$ submodules, for which each submodule $mn$ has a submap $\mathsf{M}_{mn}$ with $M_{mn}$ submodules, and~so on (Figure~\ref{fig:codinglevels}). In each submodule $mn\ldots o$ at the finest level, the code word use rate for exiting a module is
\begin{equation}
q_{mn\ldots o\curvearrowright} = \sum_{\substack{i \in \mathsf{M}_{mn\ldots o}\\ j \notin \mathsf{M}_{mn\ldots o}}}\pi_iP_{ij}\; \text{(Figure~\ref{fig:codinglevels}a)}, \label{eq:exitmno}
\end{equation}
and the total code word use rate for also visiting nodes in a module is
\begin{equation}
p^{\circlearrowright}_{mn\ldots o} = q_{mn\ldots o\curvearrowright} + \sum_{i \in \mathsf{M}_{mn\ldots o}}\pi_i. \label{eq:usemno}
\end{equation}
{such that the average code word length is}
\begin{equation}
\vspace{-6pt}
H(\mathcal{P}_{mn\ldots o}) = -\frac{q_{mn\ldots o\curvearrowright}}{p^{\circlearrowright}_{mn\ldots o}}\log{\frac{q_{mn\ldots o\curvearrowright}}{p^{\circlearrowright}_{mn\ldots o}}} - \sum_{i \in \mathsf{M}_{mn\ldots o}}\frac{\pi_i}{p^{\circlearrowright}_{mn\ldots o}}\log{\frac{\pi_i}{p^{\circlearrowright}_{mn\ldots o}}}.\label{eq:Hmno}
\end{equation}

Weighting the average code word length of the code book for module $mn\ldots o$ at the finest level by its use rate gives the contribution to the description length,
\begin{equation}
L(\mathsf{M}_{mn\ldots o}) = p^{\circlearrowright}_{mn\ldots o}H(\mathcal{P}_{mn\ldots o}).\label{eq:Lfinest}
\end{equation}

In each submodule $m$ at intermediate levels, the code word use rate for exiting to a coarser level is
\begin{equation}
q_{m\curvearrowright} = \sum_{\substack{i \in \mathsf{M}_{m}\\ j \notin \mathsf{M}_{m}}}\pi_iP_{ij},\label{eq:exitmrate}
\end{equation}
{and for entering the $M_{m}$ submodules $\mathsf{M}_{mn}$ at a finer level is}
\begin{equation}
q_{mn\curvearrowleft} = \sum_{\substack{i \notin \mathsf{M}_{mn}\\ j \in \mathsf{M}_{mn}}}\pi_iP_{ij}\;\text{(Figure~\ref{fig:codinglevels}c)}.\label{eq:entermnrate}
\end{equation}
{Therefore, the total code rate use in submodule $m$ is}
\begin{equation}
q^{\circlearrowright}_{m} = q_{m\curvearrowright} + \sum_{n=1}^{M_m}q_{mn\curvearrowleft},
\end{equation}
{which gives the average code word length}
\begin{equation}
H(\mathcal{Q}_{m}) = -\frac{q_{m\curvearrowright}}{q^{\circlearrowright}_{m}}
\log{\frac{q_{m\curvearrowright}}{q^{\circlearrowright}_{m}}}
-\sum_{n=1}^{M_m}\frac{q_{mn\curvearrowleft}}{q^{\circlearrowright}_{m}}
\log{\frac{q_{mn\curvearrowleft}}{q^{\circlearrowright}_{m}}}.
\end{equation}

Weighting the average code word length of the code book for module $m$ at intermediate levels by~its use rate, and adding the description lengths of submodules at finer levels in a recursive fashion down to the finest level in Equation~(\ref{eq:Lfinest}), gives the contribution to the description length,
\begin{equation}
	L(\mathsf{M}_m) = q^{\circlearrowright}_m H(\mathcal{Q}_m) + \sum_{n=1}^{M_m}L(\mathsf{M}_{mn}).\label{eq:Lintermediate}
	\end{equation}
	
{At the coarsest level, there is no coarser level to exit to, and the code word use rate for entering the $M$ submodules $\mathsf{M}_{mn}$ at a finer level is}
\begin{equation}
q_{m\curvearrowleft} = \sum_{\substack{i \notin \mathsf{M}_{m}\\ j \in \mathsf{M}_{m}}}\pi_iP_{ij}\; \text{(Figure~\ref{fig:codinglevels}d)},
\end{equation}
{such that the total code rate use at the coarsest level is}
\begin{equation}
q_{\curvearrowleft} =  \sum_{m=1}^{M}q_{m\curvearrowleft},
\end{equation}
\newpage{which gives the average code word length}
\begin{equation}
H(\mathcal{Q}) = 
-\sum_{m=1}^{M}\frac{q_{m\curvearrowleft}}{q_{\curvearrowleft}}
\log{\frac{q_{m\curvearrowleft}}{q_{\curvearrowleft}}}.
\end{equation}

Weighting the average code word length of the code book at the coarsest level by its use rate, and~adding the description lengths of submodules at finer levels from Equation~(\ref{eq:Lintermediate}) in a recursive fashion, gives the multilevel map equation~\cite{rosvall2011multilevel}
\begin{equation}
	L(\mathsf{M}) = q_{\curvearrowleft} H(\mathcal{Q}) + \sum_{m=1}^{M}L(\mathsf{M}_m).\label{eq:requirsivemapeq}
	\end{equation}

To find the multilevel map that best represents flows in a network, we seek the multilevel clustering of the network that minimizes the multilevel map equation over all possible multilevel clusterings of the network (see Algorithm~\ref{alg:multilevel} in Section~\ref{sec:Infomap}).

While there are several advantages with the multilevel description, including potentially better compression and effectively eliminated resolution limit~\cite{kawamoto2015estimating}, for simplicity researchers often choose two-level descriptions. In this case, there are no intermediate submodules and the two-level map equation is
\begin{equation}
L(\mathsf{M}) = q_{\curvearrowleft} H(\mathcal{Q}) + \sum_{m=1}^{M}p^{\circlearrowright}_m H(\mathcal{P}_m).\label{eq:twolevelmapeq}
\end{equation}

\subsection{The Map Equation for Higher-Order Network Flows}
The map equation for first-order network flows measures the description length of a random walker stepping between physical nodes within and between modules. This principle remains the~same also for higher-order network flows, although higher-order models guide the random walker between~physical nodes with the help of state nodes. Therefore, extending the map equation to~higher-order network flows, including those described by memory, multilayer, and sparse memory networks, is straightforward. Equations (\ref{eq:exitmno}) to (\ref{eq:twolevelmapeq}) remain the same with $i \to \alpha_i$ and $j \to \beta_j$. The~only difference is at the finest level (Figure~\ref{fig:codinglevels}b). State nodes of the same physical node assigned to the same module should share code word, or they would not represent the same object. That is, if multiple state nodes $\beta_j$ of the same physical node $i$ are assigned to the same module $mn\ldots o$, we first sum their probabilities to obtain the visit rate of physical node $i$ in module $mn\ldots o$,
\begin{equation}
\pi_{i \cap mn\ldots o} = \sum_{\beta_j \in i \cap \mathsf{M}_{mn\ldots o}}\pi_{\beta_j}\; \text{(Figure~\ref{fig:codinglevels}b)}.\label{eq:physpi}
\end{equation}

{In this way, the frequency weighted average code word length in submodule codebook $mn\ldots o$ in Equation~(\ref{eq:Hmno}) becomes}
\begin{equation}
H(\mathcal{P}_{mn\ldots o}) = -\frac{q_{mn\ldots o\curvearrowright}}{p^{\circlearrowright}_{mn\ldots o}}\log{\frac{q_{mn\ldots o\curvearrowright}}{p^{\circlearrowright}_{mn\ldots o}}} - \sum_{i \in \mathsf{M}_{mn\ldots o}}\frac{\pi_{i \cap mn\ldots o} }{p^{\circlearrowright}_{mn\ldots o}}\log{\frac{\pi_{i \cap mn\ldots o}}{p^{\circlearrowright}_{mn\ldots o}}},\label{eq:sparsememorynetworkHmno}
\end{equation}
where the sum is over all physical nodes that have state nodes assigned to module $\mathsf{M}_{mn\ldots o}$. In this way, the map equation can measure the modular description length of state-node-guided higher-order flows between physical nodes.

\begin{figure}[H]
    \centering
	\includegraphics[width=0.95\textwidth]{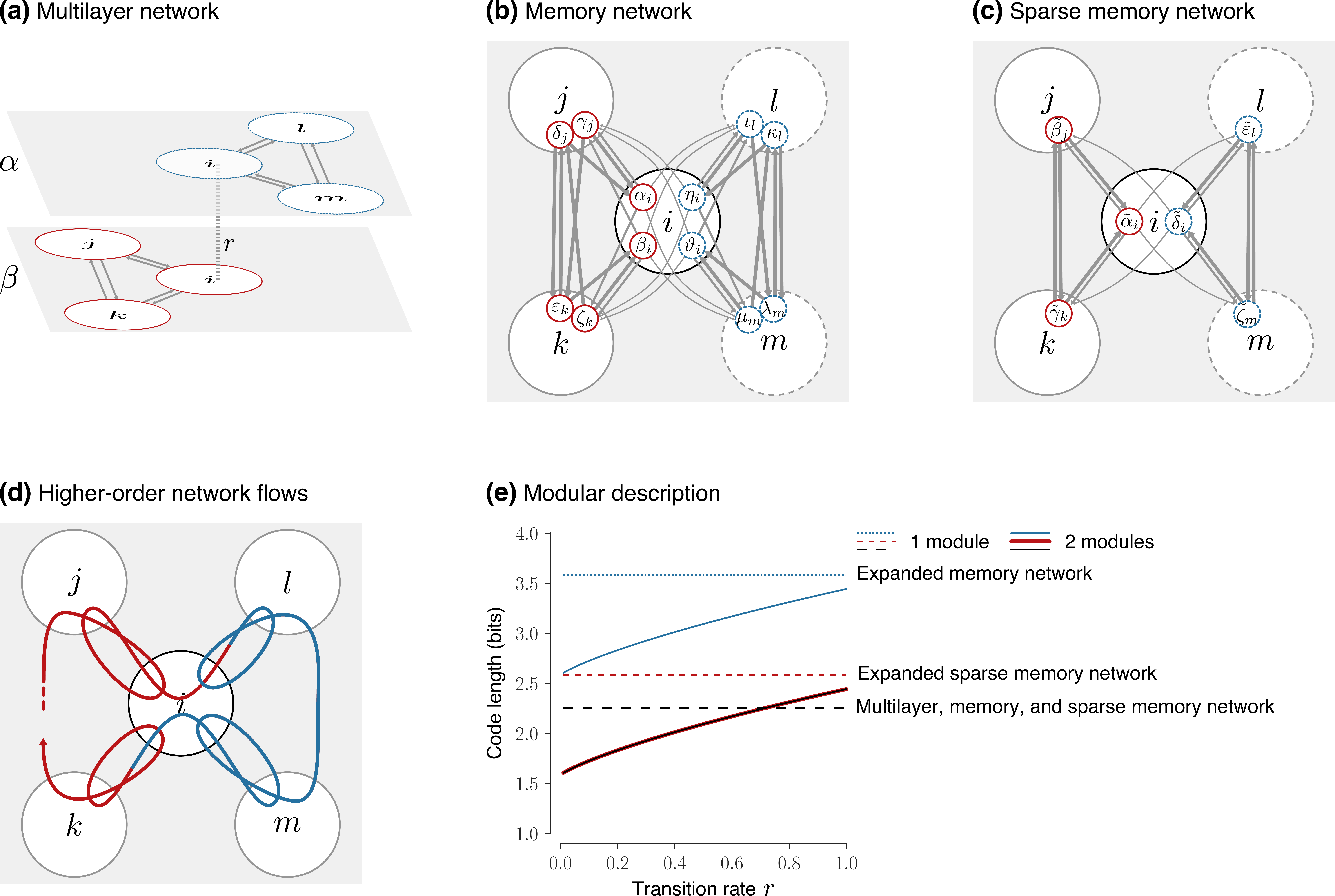}
    \caption{Mapping higher-order network flows. (\textbf{a}) A clustered multilayer network with relax rate $r$; (\textbf{b}) The same network flows in a clustered second-order memory network; (\textbf{c}) The same network flows in a clustered sparse memory network; (\textbf{d}) Sample of corresponding higher-order flow pathways; (\textbf{e}) Code length as a function of the transition rate for all representations as well as extended networks with virtual physical nodes. \label{fig:memorycoding_representation}}%
\end{figure}

To illustrate that the separation between physical nodes and state nodes matters when clustering higher-order network flows, we cluster the red state nodes and the dashed blue state nodes in Figure~\ref{fig:memorycoding_representation} in two different modules overlapping in the center physical node. For a more illustrative example, we also allow transitions between red and blue nodes at rate $r$/$2$ in the center physical node. In~the~memory and sparse memory networks, the transitions correspond to links from state nodes in physical node $i$ to~state nodes in the other module with relative weight $r$/$2$ and to the same module with relative weight $1-r$/$2$ (Figurs~\ref{fig:memorycoding_representation}b,c). In the multilayer network, the transitions correspond to relax rate $r$ according to Equation~(\ref{eq:interlayer_transition}), since relaxing to any layer in physical node $i$ with equal link weights in~both layers means that one half of relaxed flows switch layer (Figure~\ref{fig:memorycoding_representation}a). Independent of the relax rate, in~these symmetric networks the node visit rates are uniformly distributed: $\pi_{\alpha_i}=1/6$ for each of~the~six state nodes in the multilayer and sparse memory networks, and $\pi_{\alpha_i}=1/12$ for each of~the~twelve state nodes in the memory network. For illustration, if we incorrectly treat state nodes as physical nodes in the map equation, Equations (\ref{eq:exitmno}) to (\ref{eq:twolevelmapeq}), the one-module clustering of the memory network with twelve virtual physical nodes, $\mathsf{M}_{1}^{12\text{p}}$, has code length
\begin{equation}
L(\mathsf{M}_{1}^{12\text{p}}) = \underbrace{\frac{12}{12}}_{p^{\circlearrowright}_{1}}\underbrace{H\left(\frac{1}{12},\frac{1}{12},\frac{1}{12},\frac{1}{12},\frac{1}{12},\frac{1}{12},\frac{1}{12},\frac{1}{12},\frac{1}{12},\frac{1}{12},\frac{1}{12},\frac{1}{12}\right)}_{H(\mathcal{P}_1)\text{ with } H(p_1,p_2,\ldots) = -\sum_i\frac{p_i}{\sum_jp_j}\log{\frac{p_i}{\sum_jp_j}}} \approx 3.58\text{ bits.} \label{eq:mapeqexM1p12}
\end{equation}

The corresponding two-module clustering of the memory network with twelve virtual physical nodes, $\mathsf{M}_{2}^{12\text{p}}$, has code length
\begin{myequation}
L(\mathsf{M}_{2}^{12\text{p}}) = \underbrace{\frac{r}{6}}_{q_{\curvearrowleft} }\underbrace{H\left(\frac{r}{12},\frac{r}{12}\right)}_{H(\mathcal{Q})}
+2
\underbrace{\frac{6+r}{12}}_{p^{\circlearrowright}_{m}}
\underbrace{H\left(\frac{1}{12},\frac{1}{12},\frac{1}{12},\frac{1}{12},\frac{1}{12},\frac{1}{12},\frac{r}{12}\right)}_{H(\mathcal{P}_m)} \approx [2.58,3.44] \text{ for } r \in [0,1]. \label{eq:mapeqexM2p12}
\end{myequation}

Therefore, the two-module clustering gives best compression for all relax rates (Figure~\ref{fig:memorycoding_representation}e).
For~the~sparse memory network with six virtual physical nodes, the one-module clustering, $\mathsf{M}_{1}^{6\text{p}}$, has code length
\begin{equation}
L(\mathsf{M}_{1}^{6\text{p}}) = \frac{6}{6}H\left(\frac{1}{6},\frac{1}{6},\frac{1}{6},\frac{1}{6},\frac{1}{6},\frac{1}{6}\right) \approx 2.58,
\end{equation}
{and the two-module clustering, $\mathsf{M}_{2}^{6\text{p}}$, has code length}
\begin{equation}
L(\mathsf{M}_{2}^{6\text{p}}) = \frac{r}{6}H\left(\frac{r}{12},\frac{r}{12}\right)+2\frac{6+r}{12}H\left(\frac{1}{6},\frac{1}{6},\frac{1}{6},\frac{r}{12}\right) \approx [1.58, 2.44] \text{ for } r \in [0,1].
\end{equation}

While the two-module clustering again gives best compression for all relax rates, the code lengths are shifted by 1 bit compared with the memory network with twelve virtual physical nodes (Figure~\ref{fig:memorycoding_representation}e). Same dynamics but different code length. These expanded solutions with virtual physical nodes do not capture the important and special role that physical nodes play as representatives of a system's~objects.\endgraf
Properly separating state nodes and physical nodes as in Equation (\ref{eq:sparsememorynetworkHmno}) instead gives equal code lengths for identical clusterings irrespective of representation. For example, the one-module clustering of the memory network with twelve state nodes, $\mathsf{M}_{1}^{5\text{p}12\text{s}}$, and the sparse memory network with six state nodes, $\mathsf{M}_{1}^{5\text{p}6\text{s}}$, have identical code length
\begin{equation}
L(\mathsf{M}_{1}^{5\text{p}12\text{s}}) = L(\mathsf{M}_{1}^{5\text{p}6\text{s}}) = H\left(\frac{1}{6},\frac{1}{6},\frac{1}{6},\frac{1}{6},\frac{2}{6}\right) \approx 2.25,
\end{equation}
{because visits to a physical node's state nodes in the same module are aggregated according to~\mbox{Equation (\ref{eq:physpi})} such that those state nodes share code words and the encoding represents higher-order flows between a system's objects. Similarly, the two-module clustering of the memory network with~twelve state nodes, $\mathsf{M}_{2}^{5\text{p}12\text{s}}$, and the sparse memory network with six state nodes, $\mathsf{M}_{2}^{5\text{p}6\text{s}}$, have identical code length}
\begin{equation}
L(\mathsf{M}_{2}^{5\text{p}12\text{s}}) = L(\mathsf{M}_{2}^{5\text{p}6\text{s}}) = \frac{r}{6}H\left(\frac{r}{12},\frac{r}{12}\right)+2\frac{6+r}{12}H\left(\frac{1}{6},\frac{1}{6},\frac{1}{6},\frac{r}{12}\right) \approx [1.58, 2.44] \text{ for } r \in [0,1], \label{eq:mapeqex-higher-order}
\end{equation}

That is, same dynamics give same code length for identical clusterings with proper separation of~state nodes and physical nodes. 

For these solutions with physical nodes and state nodes that properly capture higher-order network flows, the overlapping two-module clustering gives best compression with relax rate \mbox{$r \lessapprox 0.71$} (Figure~\ref{fig:memorycoding_representation}e). In this example, the one-module clustering can for sufficiently high relax rate better compress the network flows than the two-module clustering. Compared with the expanded clusterings with virtual physical nodes where this cannot happen, the one-module clustering gives a~relatively shorter code length thanks to code word sharing between state nodes in physical node $i$. In~general, modeling higher-order network flows with physical nodes and state nodes, and accounting for~them when mapping the network flows, gives overlapping modular clusterings that do not depend on~the~particular representation but only on the actual dynamical patterns.

\section{Infomap}
\label{sec:Infomap}
To find the best modular description of network flows according to the map equation, we have developed the stochastic and fast search algorithm called Infomap. Infomap can operate on~standard, multilayer, memory, and sparse memory networks with  unweighted or weighted and~undirected or~directed links, and identify two-level or multilevel and non-overlapping or~overlapping clusterings. Infomap takes advantage of parallelization with OpenMP and there is also a distributed version implemented with GraphLab PowerGraph for standard networks~\cite{bae2015gossipmap}. In~principle, the search algorithm can optimize other objectives than the map equation to find other types of community~structures.

Recent versions of Infomap operate on physical nodes and state nodes. Each state node has a~unique state id and is assigned to a physical id, which can be the same for many state nodes. Infomap only uses physical nodes and state nodes for higher-order network flow representations, such as~multilayer, memory, and sparse memory networks, and physical nodes alone when they are sufficient to represent first-order network flows in standard networks.

To balance accuracy and speed, Infomap uses some repeatedly and recursively applied stochastic subroutine algorithms. For example, the multilevel clustering (Algorithm~\ref{alg:multilevel}) and two-level clustering (Algorithm~\ref{alg:two-level})
algorithms first repeatedly aggregate nodes in modules (Algorithm~\ref{alg:repeated-node-aggregation}) to optimize the map equation 
(Algorithm~\ref{alg:optimize}) and then repeatedly and recursively fine-tune (Algorithm~\ref{alg:fine-tune}) and coarse-tune (Algorithm~\ref{alg:coarse-tune}) the results. Together with complete restarts, these tuning algorithms avoid local minima and improve accuracy.

By default, Infomap tries to find a multilevel clustering of a network (Algorithm~\ref{alg:multilevel}).

\begin{algorithm}[H]
  \caption{Multilevel clustering}
  \label{alg:multilevel}
  {\fontsize{10}{16}\selectfont
  \begin{algorithmic}[1]
    \Function{multilevelPartition}{Network G}
      \State $\mathsf{M} \gets$ \Call{partition}{$G$} (Algorithm~\ref{alg:two-level})
      \State $\mathsf{M} \gets$ \Call{repeatedSuperModules}{$G$, $\mathsf{M}$} (Algorithm~\ref{alg:super})
      \State $\mathsf{M} \gets$ \Call{repeatedSubModules}{$G$, $\mathsf{M}$} (Algorithm~\ref{alg:recursive})
      \State \Return $\mathsf{M}$ \strut
  	\EndFunction
  \end{algorithmic}}
\end{algorithm}
\unskip

\begin{algorithm}[H]
  \caption{Two-level clustering}
  \label{alg:two-level}
  {\fontsize{10}{16}\selectfont
  \begin{algorithmic}[1]
    \Function{partition}{Network G}
      \State Let $L$ be the current code length
      \State Let $\epsilon$ be a small threshold \vspace{2pt}
      \State Let $I^T_{max}$ be a maximum number of iterations \vspace{2pt}
      \State $\mathsf{M} =$ \Call{repeatedNodeAggregation}{G} (Algorithm~\ref{alg:repeated-node-aggregation})
      \State $L_{old} \gets L(\mathsf{M})$ \vspace{3pt}
      \State $I^T \gets 0$
      \Repeat \vspace{2pt}
        \State $I^T \gets I^T + 1$
        \If{$I$ is odd}
        	\State $\mathsf{M} =$ \Call{FineTune}{G,M} (Algorithm~\ref{alg:fine-tune})
        \Else
        	\State $\mathsf{M} =$ \Call{CoarseTune}{G,M} (Algorithm~\ref{alg:coarse-tune}) \vspace{2pt}
        \EndIf
        \State $\delta L \gets L(\mathsf{M}) - L_{old}$
        \State $L_{old} \gets L(\mathsf{M})$ \vspace{2pt}
      \Until{$\delta L > -\epsilon$ \OR $I^T = I^T_{max}$} \strut
     \EndFunction
  \end{algorithmic}}
\end{algorithm}

\subsection{Two-Level Clustering}
The complete two-level clustering algorithm improves the repeated node aggregation algorithm (Algorithm~\ref{alg:repeated-node-aggregation}) by breaking up modules of sub-optimally aggregated nodes. That is, once the~algorithm assigns nodes to the same module, they are forced to move jointly when Infomap rebuilds the~network, and what was an optimal move early in the algorithm might have the opposite effect later in~the~algorithm.
Similarly, two or more modules that merge together and form a single module when the algorithm rebuilds the network can never be separated again in the repeated node aggregation algorithm.
Therefore, the accuracy can be improved by breaking up modules of~the~final state of~the~repeated node aggregation algorithm and trying to move individual nodes or smaller submodules into new or neighboring modules. The two-level clustering algorithm (Algorithm~\ref{alg:two-level}) performs this refinement by iterating fine-tuning (Algorithm~\ref{alg:fine-tune}) and coarse-tuning (Algorithm~\ref{alg:coarse-tune}).

\subsection{Repeated Node Aggregation}

The repeated node aggregation in Algorithm~\ref{alg:repeated-node-aggregation} follows closely the machinery of the Louvain method~\cite{blondel2008fast}. While the Louvain method seeks to maximize the modularity score, which fundamentally operates on link density rather than network flows, its repeated node aggregation machinery is effective also for optimizing other objectives than the modularity score. 

Infomap aggregates neighboring nodes into modules, subsequently aggregates them into larger modules, and repeats. First, Infomap assigns each node to its own module. Then, in random sequential order, it tries to move each node to the neighboring or a new module that gives the largest code length decrease (Algorithm~\ref{alg:optimize}). If no move gives a code length decrease, the node stays in its original module. The change of code length for each possible move can be calculated locally based only on the change in $enter$ and $exit$ flows, the current flow in the moving node and affected modules, and the initial $enter$ flow for the module. For state nodes, the physical node visit rates can be locally updated.

Infomap repeats this procedure, each time in a new random sequential order, until no move happens. Then Infomap rebuilds the network with the modules of the last level forming the nodes at~the~new level, and, exactly as in the previous level, aggregates the nodes into modules, which replace the existing ones. Infomap repeats this hierarchical network rebuilding until the map equation cannot be further reduced.

The computational complexity of the repeated node aggregation is considered to be $\mathcal{O}(N \log N)$ for the Louvain method applied to networks with $N$ nodes~\cite{blondel2008fast}. While Infomap optimizes the map equation with some more costly logarithms, there is no fundamental difference in the scaling. However, for sparse memory networks, the scaling applies to the number of state nodes. In any case, the first aggregation step is the most expensive because Infomap considers each node and~all its links such that the complexity scales with the number of~links in the network. To further optimize the map equation, Infomap performs stochastic tuning steps with more network dependent computational complexity.

\begin{algorithm}[H]
  \caption{Repeated node aggregation}
  \label{alg:repeated-node-aggregation}
  \begin{algorithmic}[1]
    \Function{repeatedNodeAggregation}{Network $G$, Node-to-module map $\mathsf{M}$}
      \State Let $L = L(\mathsf{M})$ be the code length for the map $\mathsf{M}$ with node-to-module assignments
      \State Let $\epsilon$ be a small threshold \vspace{2pt}
      \State Let $I^L_{max}$ be a maximum number of level iterations \vspace{2pt}
      \If{$\mathsf{M}$ is empty}
      	\State $\mathsf{M} \gets$ one module per node in $G$
      \EndIf
      \State $L_{old} \gets L(\mathsf{M})$
      \State $I^L \gets 0$
      \Repeat
        \State $I^L \gets I^L + 1$
        \State $N \gets $ number of nodes in $G$
        \State $\mathsf{M} =$ \Call{optimize}{$G$, $\mathsf{M}$} (Algorithm~\ref{alg:optimize})
        \State $M \gets $ number of modules in $\mathsf{M}$
        \State $L = L(\mathsf{M})$
        \State $\delta L \gets L - L_{old}$
        \State $L_{old} \gets L$ \vspace{2pt}
        \If{$M = 1$ \OR $M = N$}
          \State Break \strut
        \EndIf
        \State $G \gets $ Aggregate nodes within modules and links between modules to a new network
      	\State $\mathsf{M} \gets$ one module per node in $G$ \vspace{2pt}
      \Until{$\delta L > -\epsilon$ \OR $I^L = I^L_{max}$}
      \State \Return $\mathsf{M}$ \strut
    \EndFunction
  \end{algorithmic}
\end{algorithm}
\unskip

\begin{algorithm}[H]
  \caption{Optimize network}
  \label{alg:optimize}
  \begin{algorithmic}[1]
  	\Function{optimize}{Network $G$, Node-to-module map $\mathsf{M}$}
      \State Let $I^M_{max}$ be a maximum number of iterations
      \State $I^M \gets 0$
      \Repeat
      	\State $I^M \gets I^M + 1$
        \For{node $n_i \in G$ in random order}
          \For{link $l_{ij}$ to neighboring node $n_j$ in module $m_j$}
            \parState{Calculate change in $enter$ and $exit$ flows for modules $m_i$ and $m_j$ if moving node $n_i$ to module $m_j$}
            \parState{For state nodes, calculate change in physical node visit rates (Equation~(\ref{eq:physpi}))}\strut
          \EndFor
          \State Calculate change in $enter$ and $exit$ flow if moving to a new module
          \parState{Calculate change in code length $\delta L_{ij}$ for each possible move of node $n_i$ to module $m_j$}
          \If{$\min_{j} \delta L_{ij} < -\epsilon$ \AND $m_j \neq m_i$}
              \State Update $\mathsf{M}$ with $m_i \gets m_j$
              \parState{Update $enter$, $exit$ and node flow in modules $m_i$ and $m_j$}
              \State $L \gets L + \delta L_{ij}$ \strut \vspace{3pt}
          \EndIf
        \EndFor
        \State $\delta L \gets L - L_{old}$ \vspace{2pt}
        \State $L_{old} \gets L$ \vspace{2pt}
      \Until{$\delta L > -\epsilon$ \OR $I^M = I^M_{max}$} \vspace{2pt}
      \State \Return $\mathsf{M}$ \strut
    \EndFunction
  \end{algorithmic}
\end{algorithm}

\subsection{Fine-Tuning}
\label{subsec:fine-tune}
In the fine-tuning step, Infomap tries to move single nodes out from existing modules into new or~neighboring modules (Algorithm~\ref{alg:fine-tune}). First, Infomap reassigns each node to be the sole member of its own module to allow for single-node movements. Then it moves all nodes back to their respective modules of the previous step. At this stage, with the same clustering as in the previous step but with each node free to move between the modules, Infomap reapplies the repeated node aggregation (Algorithm~\ref{alg:repeated-node-aggregation}).
\begin{algorithm}[H]
  \caption{Fine-tuning}
  \label{alg:fine-tune}
  \begin{algorithmic}[1]
    \Function{FineTune}{Network $G$, Node-to-module map $\mathsf{M}$}
      \State $\widetilde{\mathsf{M}} \gets$ one module per node in $G$
      \State Update $\widetilde{\mathsf{M}}$ to mirror $\mathsf{M}$
      \State $\mathsf{M} \gets$ \Call{repeatedNodeAggregation}{$G$, $\widetilde{\mathsf{M}}$} (Algorithm~\ref{alg:repeated-node-aggregation}).
      \State \Return $\mathsf{M}$ \strut
    \EndFunction
  \end{algorithmic}
\end{algorithm}

\subsection{Coarse-Tuning}
\label{subsec:coarse-tune}
In the course-tuning step, Infomap tries to move submodules out from existing modules into~new or neighboring modules (Algorithm~\ref{alg:coarse-tune}).
First, Infomap treats each module as a network on its own and~applies repeated node aggregation (Algorithm~\ref{alg:repeated-node-aggregation}) on this network. This procedure generates one or more submodules for each module. Infomap then replaces the modules by the submodules and~moves the submodules into their modules as an initial solution. At this stage, with the same clustering as~in~the~previous step but with each submodule free to move between the modules, Infomap reapplies the repeated node aggregation (Algorithm~\ref{alg:repeated-node-aggregation}).

\subsection{Multilevel Clustering}
Infomap identifies multilevel clusterings by extending two-level clusterings both by iteratively finding supermodules of modules and by recursively finding submodules of modules. To find the~optimal multilevel solution, Infomap first tries to find an optimal two-level clustering, then iteratively tries to find supermodules and then recursively tries to cluster those modules until the code length cannot be further improved (Algorithm~\ref{alg:multilevel}).

\begin{algorithm}[H]
  \caption{Coarse-tuning}
  \label{alg:coarse-tune}
  \begin{algorithmic}[1]
    \Function{CoarseTune}{Network $G$, Optional node-to-module map $\mathsf{M}$}
      \For{module $m_i$ in $\mathsf{M}$}
      	\State $G_i \gets $ network of nodes and links within $m_i$
      	\State Sub-map $\mathsf{M}_i \gets $ \Call{repeatedNodeAggregation}{$G_i$} (Algorithm~\ref{alg:repeated-node-aggregation}) \strut
      \EndFor
      \State $\widetilde{G} \gets $ network of submodules $\mathsf{M}_i$
      \State $\widetilde{\mathsf{M}} \gets$ map of nodes in $\widetilde{G}$ to $\mathsf{M}$
      \State $\widetilde{\mathsf{M}} \gets$ \Call{repeatedNodeAggregation}{$\widetilde{G}$, $\widetilde{\mathsf{M}}$} (Algorithm~\ref{alg:repeated-node-aggregation}).
      \State $\mathsf{M} \gets \mathsf{M}(G)$ such that $\mathsf{M}(G) \Leftrightarrow \widetilde{\mathsf{M}}(\widetilde{G})$
      \State \Return $\mathsf{M}$ \strut
    \EndFunction
  \end{algorithmic}
\end{algorithm}

To identify a hierarchy of supermodules from a two-level clustering, Infomap first tries to~find a~shorter description of flows between modules. It iteratively runs the two-level clustering algorithm (Algorithm~\ref{alg:two-level}) on a network with one node for each module at the coarsest level, node-visit rates from~the~module-entry rates, and links that describe aggregated flows between the modules (Algorithm~\ref{alg:super}). For each such step, a two-level code book replaces the coarsest code book. In this way, describing entries into, within, and out from supermodules at a new coarsest level replaces only describing entries into the previously coarsest modules. 

\begin{algorithm}[H]
  \caption{Repeated supermodule clustering}
  \label{alg:super}
  \begin{algorithmic}[1]
    \Function{repeatedSuperModules}{Network $G$, Node-to-module map $\mathsf{M}$}
      \State Let $L = L(\mathsf{M})$ be the code length for the map $\mathsf{M}$ with node-to-module assignments
      \State Let $\widehat{L}(\mathsf{M})$ be the index code length of $\mathsf{M}$
      \State Let $\epsilon$ be a small threshold
      \State $\widehat{L}_{old} \gets \widehat{L}(\mathsf{M})$
      \Repeat
	    \parState{$\widehat{G} \gets$ A new network with one node for each module at the coarsest level of $\mathsf{M}$, node-visit rates from the module-entry rates, and links that describe aggregated flows between the modules \label{alg:super:1}} \strut \vspace{-12pt}
        \State $\widehat{N} \gets $ number of nodes in $\widehat{G}$
        \State $\widehat{\mathsf{M}} \gets$ \Call{partition}{$\widehat{G}$} (Algorithm~\ref{alg:two-level})
        \State $\widehat{M} \gets $ number of modules in $\widehat{\mathsf{M}}$
        \State $\delta L \gets L(\widehat{\mathsf{M}}) - \widehat{L}_{old}$
        \State $\widehat{L}_{old} \gets \widehat{L}(\widehat{\mathsf{M}})$ \strut
        \If{$\widehat{M} = 1$ \OR $\widehat{M} = \widehat{N}$}
          \State Break \strut
        \EndIf
        \State $\mathsf{M} \gets$ A multilevel map composed of $\widehat{\mathsf{M}}$ and $\mathsf{M}$
      \Until{$\delta L > -\epsilon$}
      \State \Return $\mathsf{M}$ \strut
    \EndFunction
  \end{algorithmic}
\end{algorithm}

For a fast multilevel clustering, Infomap can keep this hierarchy of supermodules and try to~recursively cluster the bottom modules into submodules with Algorithm~\ref{alg:recursive}. However, as this hierarchy of supermodules is constrained by the initial optimal two-level clustering, discarding everything but~the~top supermodules and starting the recursive submodule clustering from them often identifies a~better multilevel clustering. To find submodules, Infomap runs the two-level clustering (Algorithm~\ref{alg:two-level}) followed by the super-module clustering (Algorithm \ref{alg:super}) algorithm on the network within each module to find the largest submodule structures. If Infomap finds non-trivial submodules (that is, not one submodule for all nodes and not one submodule for each node), it replaces the module with~the~submodules and collects the submodules in a queue to explore each submodule in parallel for~each hierarchical level. In this way, Infomap can explore the submodule hierarchy in parallel with~a~breadth-first search to efficiently find a multilevel clustering of the input network.

\begin{algorithm}[H]
  \caption{Repeated submodule clustering}
  \label{alg:recursive}
  \begin{algorithmic}[1]
    \Function{repeatedSubModules}{Network $G$, multilevel node-to-module map $\mathsf{M}$}
      \State $Q \gets$ top modules in $\mathsf{M}$ 
      \State Remove submodules in $\mathsf{M}$ if exist
      \While{$Q$ not empty}
        \State $R \gets$ empty queue to collect submodules
        \For{module $m_i$ in $Q$ (in parallel)}
          \State $G_i \gets $ network of nodes and links within $m_i$
          \State $N \gets$ number of nodes in $G_i$
          \State $\mathsf{M}_i \gets $ \Call{partition}{$G_i$} (Algorithm~\ref{alg:two-level}) \strut
          \State $\mathsf{M}_i \gets $ \Call{repeatedSuperModules}{$G_i$, $\mathsf{M}_i$} (Algorithm~\ref{alg:super}) \strut
          \State $\widehat{M} \gets$ number of top modules in $\mathsf{M}_i$
          \If{$\widehat{M} = 1$ \OR $\widehat{M} = N$}
          	\State Replace $m_i$ with top modules in $\mathsf{M}_i$
            \For{each top module $\widehat{m}$ in $\mathsf{M}_i$}
              \parState{Add $\widehat{m}$ to queue $R$\\}
            \EndFor
          \EndIf
        \EndFor
        \State $Q \gets R$
      \EndWhile
      \State \Return $\mathsf{M}$ \strut
    \EndFunction
  \end{algorithmic}
\end{algorithm}

\subsection{Download Infomap}
Infomap can be downloaded from \url{www.mapequation.org}, which contains installation instructions, a complete list of running options, and examples for how to run Infomap stand-alone or with other software, including igraph, Python, Jupyter, and R. In the Appendix, we provide Infomap's basic syntax (Appendix~\ref{appendix:infomap_syntax}) and examples for how to cluster a multilayer network (Appendix~\ref{appendix:fig3a.net}), a memory network (Appendix~\ref{appendix:fig3b.net}), and a sparse memory network (Appendix~\ref{appendix:fig3c.net}). The website also contains interactive storyboards to explain the mechanics of the map equation, and interactive visualizations to simplify large-scale networks and their changes over time.

\subsection{Mapping Multistep and Multi-Source Data with Infomap}
We illustrate our approach by mapping network flows from air traffic data between more than 400 airports in the US~\cite{USairtraffic}. For quarterly reported itineraries in 2011, we assembled pathways into~four quarterly, two half-year, and one full-year collection. From pathways of length two, three, and~four, for~each collection we generated sparse memory networks corresponding to first-, second-, and~third-order Markov models of flows. Then we generated sparse memory networks corresponding to multilayer networks with four layers from the quarterly collections and two layers from the half-year collections for all Markov orders. This example illustrates how sparse memory networks can go beyond standard representations and represent combinations of multilayer and memory networks.

To identify communities in multistep and multi-source air traffic data represented with sparse memory networks, we run Infomap ten times and report the median values in Table~\ref{tab:empirical}. Except for~a~small drop in the number of physical nodes for the third-order models because some airports are not represented in long itineraries, the number of physical nodes remain the same for all models. However, the number of state nodes and links between them increases with order. Larger systems require state-node lumping based on minimum information loss into more efficient sparse memory networks that correspond to variable-order Markov models. Such more compact descriptions of~higher-order can balance over- and underfitting as well as cut the computational time~\cite{persson2016maps}. In~any case, higher-order representations can better capture constraints on flows. For example, a~drop in~the~entropy rate by one bit corresponds to doubled flow prediction, and the four-bit drop between~the~first- and third-order models corresponds to a sixteenfold higher flow prediction accuracy. These flows move with relatively long persistence times among groups of airports that form connected legs in frequent itineraries. Infomap capitalizes on this modular flow pattern and identifies pervasively overlapping modules. 

\begin{table}[H]
\caption{Mapping higher-order network flows from air traffic data. Sparse memory networks corresponding to Markov order $m=$1, 2, and 3 in one layer with data from quarters $1+2+3+4$, in two layers with data from quarters $1+2$ and $3+4$, and in four layers with data from quarters 1, 2, 3, and 4, respectively. We report the number of physical node $N_P$, state nodes $N_S$, and links $N_L$. The~entropy rate $H$ measures the number of bits required to predict where flows are going. $D$ is the node weighted average depth of nodes in the multilevel clustering. The~perplexity $P$ of the module size distribution measures the effective number of modules, and~the~physical node weighted average of the perplexity of module assignments. $A$ is the effective number of module assignments per~physical node. Finally, the time corresponds to runs without parallelization on a 3.60 GHz desktop computer. We use relax rate $r=0.25$ and all values are medians of ten runs with Infomap}
\centering
\begin{tabular}{llllrrSSSSr}
\toprule
\textbf{Quarters}	& \boldmath{$m$}	& \boldmath{$l$} & \boldmath{$N_P$}	& \boldmath{$N_S$}	& \boldmath{$N_L$}	& \boldmath{$H$} &	\boldmath{$D$}	& \boldmath{$P$}	& \boldmath{$A$}	& \textbf{Time} \\
\midrule
$1+2+3+4$ & 1 & 1 & 438 & \num{438}    & \num{9681}     & 5.1  & 2.0 & 1.0 & 1.0 & 0.016 s \\
$1+2,3+4$ & 1 & 2 & 438 & \num{861}    & \num{34384}    & 5.1  & 2.0 & 1.2 & 1.0 & 0.063 s \\
$1,2,3,4$ & 1 & 4 & 438 & \num{1683}   & \num{121749}   & 5.0  & 2.0 & 4.0 & 1.0 & 0.14 s \\[0.5ex]
$1+2+3+4$ & 2 & 1 & 438 & \num{9681}   & \num{181326}   & 3.8  & 2.4 & 13  & 3.7 & 1.7 s \\
$1+2,3+4$ & 2 & 2 & 438 & \num{17203}  & \num{614472}   & 3.8  & 3.0 & 15  & 3.8 & 10 s \\
$1,2,3,4$ & 2 & 4 & 438 & \num{30489}  & \num{2014650}  & 3.7  & 3.0 & 13  & 3.3 & 27 s \\[0.5ex]
$1+2+3+4$ & 3 & 1 & 432 & \num{180900} & \num{465456}   & 1.1  & 4.0 & 25  & 5.7 & 8.2 min \\
$1+2,3+4$ & 3 & 2 & 432 & \num{307904} & \num{1406605}  & 0.95 & 3.0 & 32  & 5.0 & 22 min \\
$1,2,3,4$ & 3 & 4 & 432 & \num{507054} & \num{4112089}  & 0.91 & 3.0 & 19  & 3.8 & 52 min \\
\bottomrule
\end{tabular}
\label{tab:empirical}
\end{table}

\section{Conclusions}
The map equation applied to sparse memory networks provides a general solution to~reveal modular patterns in higher-order flows through complex systems. Rather than multiple community-detections algorithms for a range of network representations, the map equation's search algorithm Infomap can be applied to general higher-order network flows described by sparse memory networks. A sparse memory network uses abstract state nodes to describe higher-order dynamics and~physical nodes to represent a system's objects. This distinction makes all the difference. The~flexible sparse memory network can efficiently describe higher-order network flows of various types such as flows in memory and multilayer networks or their combinations. Simplifying and highlighting important patterns in these flows with the map equation and Infomap open for more effective analysis of complex systems.

\vspace{6pt}
\acknowledgments{We are grateful to Christian Persson and Manlio De Domenico for helpful discussions. Martin~Rosvall was supported by the Swedish Research Council grant 2016-00796.}

\authorcontributions{D.E, L.B., and M.R conceived and designed the experiments; D.E. performed the experiments. D.E, L.B., and M.R analyzed the data and wrote the paper.}

\conflictsofinterest{The authors declare no conflict of interest.} 

\appendixtitles{yes} 
\appendixsections{multiple} 
\appendix
\section{Running Infomap}
\vspace{-6pt}
\subsection{Infomap Syntax}\label{appendix:infomap_syntax}
To run Infomap on a network, use the command 
\begin{small}
\begin{lstlisting}[language=bash,moredelim={[is][\textbf]{@}{@}}]
@./Infomap [options] network_data dest@
\end{lstlisting}
\end{small}

The option \lstinline|network_data| should point to a valid network file and \lstinline|dest| to a directory where Infomap should write the output files. If no option is given, Infomap will assume an~undirected network and try to cluster it hierarchically. Optional arguments, including directed or~undirected links, two-level or multilevel solutions, or various input formats, can be put anywhere. Run~\lstinline|./Infomap --help| for a list and explanation of supported arguments.

\subsection{Clustering a Multilayer Network}
\label{appendix:fig3a.net}

The multilayer network in Figure~\ref{fig:memorycoding_representation}a can be described with the multilayer network format in~\lstinline|fig3a.net| below and clustered for relax rate $r=0.4$ with the command
\begin{small}
\begin{lstlisting}[language=bash,moredelim={[is][\textbf]{@}{@}}]
@./Infomap --input-format multilayer --multilayer-relax-rate 0.4 fig3a.net .@
\end{lstlisting}
\end{small}

See Appendix~\ref{appendix:fig3c.tree} for the overlapping clustering output, and Appendix \ref{appendix:fig3c.net} for an alternative representation with a sparse memory network. In fact, Infomap internally represents the multilayer network in \lstinline|fig3a.net| for $r=0.4$ with the sparse memory network in Appendix \ref{appendix:fig3c.net} with transition rates $r$/$2$ between state nodes in different layers, since $r$/$2$ stays among state nodes in the same layer in this symmetric two-layer network.

\begin{small}
\begin{lstlisting}[label={appendix:lst:multilayer-format-example}]
# fig3a.net - Multilayer network
# Lines starting with # are ignored
*Vertices 5
#physicalId name
1 "i"
2 "j"
3 "k"
4 "l"
5 "m"
*Intra
#layerId physicalId physicalId weight
1 1 4 1
1 4 1 1
1 1 5 1
1 5 1 1
1 4 5 1
1 5 4 1
2 1 2 1
2 2 1 1
2 1 3 1
2 3 1 1
2 2 3 1
2 3 2 1
\end{lstlisting}
\end{small}

The inter-layer links will be generated based on the multilayer relax rate, but could also be modeled explicitly with an \lstinline|*Inter| section as below

\begin{small}
\begin{lstlisting}[label={appendix:lst:multilayer-inter-layer}]
*Inter
#layerId physicalId layerId weight
1 1 2 1
2 1 1 1
\end{lstlisting}
\end{small}

\subsection{Clustering a Memory Network}
\label{appendix:fig3b.net}
The memory network in Figure~\ref{fig:memorycoding_representation}b with transition rates corresponding to relax rate $r = 0.4$ for~the~multilayer network in Figure~\ref{fig:memorycoding_representation}a, can be described with the memory network format in~\lstinline|fig3b.net| below and clustered with the command
\begin{small}
\begin{lstlisting}[language=bash,moredelim={[is][\textbf]{@}{@}}]
@./Infomap --input-format memory fig3b.net .@
\end{lstlisting}
\end{small}

See Appendix~\ref{appendix:fig3c.tree} for the overlapping clustering output, and Appendix \ref{appendix:fig3c.net} for an alternative representation with a sparse memory network.

\begin{small}
\begin{lstlisting}[label={appendix:lst:3gram-format}]
# fig3b.net - Memory network
# Lines starting with # are ignored
*Vertices 5
#physicalId name
1 "i"
2 "j"
3 "k"
4 "l"
5 "m"
*3grams
#from through to weight
2 1 2 0.8
2 1 3 0.8
2 1 4 0.2
2 1 5 0.2
3 1 2 0.8
3 1 3 0.8
3 1 4 0.2
3 1 5 0.2
1 2 1 1
1 2 3 1
3 2 1 1
3 2 3 1
2 3 1 1
2 3 2 1
1 3 1 1
1 3 2 1
4 1 4 0.8
4 1 5 0.8
4 1 2 0.2
4 1 3 0.2
5 1 4 0.8
5 1 5 0.8
5 1 2 0.2
5 1 3 0.2
1 4 1 1
1 4 5 1
5 4 1 1
5 4 5 1
1 5 1 1
1 5 4 1
4 5 1 1
4 5 4 1
\end{lstlisting}
\end{small}

\subsection{Clustering a Sparse Memory Network}
The sparse memory network in Figure~\ref{fig:memorycoding_representation}c with transition rates corresponding to relax rate $r = 0.4$ for the multilayer network in Figure~\ref{fig:memorycoding_representation}a, can be described with the sparse memory network format in \lstinline|fig3c.net| below and clustered with the command
\begin{small}
\begin{lstlisting}[language=bash,moredelim={[is][\textbf]{@}{@}}]
@./Infomap --input-format sparse fig3c.net.@
\end{lstlisting}
\end{small}
See Appendix~\ref{appendix:fig3c.tree} for the overlapping clustering output.

\label{appendix:fig3c.net}
\begin{small}
\begin{lstlisting}[label={appendix:lst:state-format-example}]
# fig3c.net - Sparse memory network
# Lines starting with # are ignored
*Vertices 5
#physicalId name
1 "i"
2 "j"
3 "k"
4 "l"
5 "m"
*States
#stateId physicalId name
1 1 "alpha~_i"
2 2 "beta~_j"
3 3 "gamma~_k"
4 1 "delta~_i"
5 4 "epsilon~_l"
6 5 "zeta~_m"
*Links
#sourceStateId targetStateId weight
1 2 0.8
1 3 0.8
1 5 0.2
1 6 0.2
2 1 1
2 3 1
3 1 1
3 2 1
4 5 0.8
4 6 0.8
4 2 0.2
4 3 0.2
5 4 1
5 6 1
6 4 1
6 5 1
\end{lstlisting}
\end{small}

For reference, the memory network in Appendix \ref{appendix:fig3b.net} can also be represented with the sparse memory network format in \lstinline|fig3b_sparse.net| below and clustered with the command
\begin{small}
\begin{lstlisting}[language=bash,moredelim={[is][\textbf]{@}{@}}]
@./Infomap --input-format sparse fig3b_sparse.net .@
\end{lstlisting}
\end{small}
See Appendix~\ref{appendix:fig3c.tree} for the overlapping clustering output.

\begin{small}
\begin{lstlisting}[label={appendix:lst:3gram-state-format}]
# fig3b_sparse.net - state network
# Lines starting with # are ignored
*Vertices 5
#physicalId name
1 "i"
2 "j"
3 "k"
4 "l"
5 "m"
*States
#stateId physicalId name
1 1 "alpha_i"
2 1 "beta_i"
3 2 "gamma_j"
4 2 "delta_j"
5 3 "epsilon_k"
6 3 "zeta_k"
7 1 "eta_i"
8 1 "theta_i"
9 4 "iota_l"
10 4 "kappa_l"
11 5 "lambda_m"
12 5 "mu_m"
*Links
#sourceStateId targetStateId weight
1 3 0.8
1 6 0.8
1 9 0.2
1 12 0.2
2 3 0.8
2 6 0.8
2 9 0.2
2 12 0.2
3 1 1
3 5 1
4 1 1
4 5 1
5 2 1
5 4 1
6 2 1
6 4 1
7 9 0.8
7 12 0.8
7 3 0.2
7 6 0.2
8 9 0.8
8 12 0.8
8 3 0.2
8 6 0.2
9 7 1
9 11 1
10 7 1
10 11 1
11 8 1
11 10 1
12 8 1
12 10 1
\end{lstlisting}
\end{small}

\subsection{Clustering Output}
\label{appendix:fig3c.tree}
All examples in Appendices~\ref{appendix:fig3a.net}--\ref{appendix:fig3c.net} give the same overlapping physical node clustering below, where \lstinline|1:1 0.166667 "i" 1| says, reading from right to left, that the physical node with id 1 is called i and has visit rate 0.166667 as the first node in the first module.
\begin{small}
\begin{lstlisting}[label={appendix:lst:tree-format}]
# A tree output from running Infomap on the example networks in Figure 3
# path flow name physicalId:
1:1 0.166667 "i" 1
1:2 0.166667 "j" 2
1:3 0.166667 "k" 3
2:1 0.166667 "i" 1
2:2 0.166667 "l" 4
2:3 0.166667 "m" 5
\end{lstlisting}
\end{small}

With the optional argument \lstinline|--expanded|, Infomap gives the clustering of each state node (not~shown). 






\end{document}